\documentstyle[epsfig]{mn}
\begin{document}

\title[Circular polarisation from GRS~1915+105]
{
Variable circular polarisation associated
with relativistic ejections from GRS 1915+105
}
\author[R.~P.~Fender  et al.]
{R. P. Fender$^1$,
D. Rayner$^2$, 
D.G. McCormick$^3$, T.W.B. Muxlow$^3$, 
G.G. Pooley$^4$, \cr R.J. Sault$^2$,
R.E. Spencer$^3$
\small
\\
$^1$ Astronomical Institute `Anton Pannekoek', University of Amsterdam,
and Center for High Energy Astrophysics, Kruislaan 403, \\
1098 SJ, Amsterdam, The Netherlands {\bf rpf@astro.uva.nl}\\
$^2$ Australia Telescope National Facility, CSIRO, PO Box 76, Epping
NSW 2121, Australia\\
$^3$ Jodrell Bank Observatory, University of Manchester, Cheshire, SK11 9DL\\
$^4$ Mullard Radio Astronomy Observatory, Cavendish Laboratory,
Madingley Road, Cambridge CB3 OHE\\
}

\maketitle

\begin{abstract}

We report the discovery of variable circularly polarised radio
emission associated with relativistic ejections from GRS 1915+105,
based on observations with the Australia Telescope Compact Array
(ATCA) and the Multi-Element Radio-Linked Interferometer Network
(MERLIN). Following a radio flare in 2001 January, significant and
variable circular polarisation, at a fractional level of 0.2--0.4\%,
was measured with ATCA at four frequencies between 1--9 GHz. Following
an additional outburst 65 days later in 2001 March, further ATCA
observations measured a comparable sign and level of circular
polarisation at two frequencies. At this second epoch, contemporaneous
MERLIN observations directly imaged a relativistic ejection event and 
allowed us to confidently associate both the circularly and linearly
polarised emission with the relativistic ejecta, allowing a
detailed measurement of the full polarisation properties in the
optically thin phase. The fractional circular polarisation spectrum
appears to flatten at higher frequencies in 2001 January, when there
is strong evidence for multiple components at different optical
depths. While we cannot conclusively distinguish between synchrotron
or propagation-induced conversion as the origin of the circularly
polarised component, we do not consider that coherent or birefringent
scintillation mechanisms are likely. The implication is therefore that
the ejections from GRS 1915+105 are associated with a significant
population of low-energy electrons, with associated consequences for the
energetics of relativistic ejection events. We briefly compare the
data for SS 433 and GRS 1915+105 with AGN and note that
linear--to--circular polarisation ratios $<1$ are observed in both
binary sources at the lowest frequency, and ratios $>1$ at the higher
frequencies, illustrating the role of Faraday depolarisation.  In
addition, the 2001 January ATCA observations reveal a linear
polarisation `rotator' event, probing the (variable) orientation of
the magnetic field structure in the outflow.

\end{abstract}


\section{Introduction}

Determining the physical composition and magnetic field structure of
relativistic ejecta from black holes in Active Galactic Nuclei (AGN)
and their local cousins, black holes and neutron stars in X-ray
binaries (XRBs), is a fundamental question for the physics and
energetics of accretion in the environment of such extreme
objects. By measuring such quantities we can probe the probable
magnetohydrodynamic formation of jets and the flow of mass at a
handful of Schwarzschild radii. 

Estimating the relative fractions of p$^+$:e$^-$ `normal' baryonic
matter and e$^+$:e$^-$ `antimatter' pairs is not a straightforward
exercise. However, in recent years there has been a revival of
interest in the possibility of probing the composition and power of
such relativistic jets by means of observing the circularly polarised
(CP) component of their radio emission (Wardle \& Homan 2002; Falcke
et al. 2002; Macquart 2002).  The CP component may arise directly from
the synchrotron process if the matter is baryonic (e.g. Legg \&
Westfold 1968), from linear polarisation (LP) $\rightarrow$ CP
`repolarisation' (e.g. Pacholczyk 1973; Jones \& O'Dell 1977; Kennett
\& Melrose 1998) in a baryonic or pair plasma, or via some other
poorly-understood mechanism.  Recent circular polarisation (CP)
observations of AGN (e.g. Wardle et al. 1998; Macquart et al. 2000;
Rayner, Norris \& Sault 2000; Homan, Attridge \& Wardle 2001) have
been used in an attempt to probe directly the composition and
distribution of the particles in the jets from these supermassive
black holes.

Closer to home, circular polarisation has been detected from the
(probable) low-luminosity AGN (LLAGN) Sgr A* at the galactic centre
(Bower, Falcke \& Backer 1999; Sault \& Marquart 1999) and the related
nearby radio core in M81 (Brunthaler et al. 2001).  In 2000 May
we discovered circularly polarised radio emission from the famous
galactic jet source SS 433 (Fender et al. 2000).  This X-ray binary was
clearly detected at four frequencies between 1--9 GHz with a CP
spectrum decreasing with increasing frequency. However, our lack of
ability to resolve the variable extended structure of SS 433 with the
Australia Telescope Compact Array (ATCA) did not enable us to
determine accurately the fractional CP spectrum and hence get a clear
grip on the physical process which might be responsible. Furthermore,
SS 433 seems to be unique amongst relativistic jet sources as it
appears to have a fixed jet velocity of $0.26c$ and displays optical /
infrared / X-ray emission lines which seem to clearly indicate a large
baryonic fraction (e.g. Margon 1984; Kotani et al. 1994).


\subsection{The `microquasar' GRS 1915+105}

The X-ray binary GRS 1915+105 was the first galactic system to display
apparent superluminal motions within our galaxy (Mirabel \& Rodriguez
1994), implying large bulk velocities maybe comparable to those in
AGN. The system has been directly resolved, on multiple occasions,
into highly relativistic jets which display different patterns of
behaviour closely linked to the X-ray state of the source (Fender et
al. 1999; Dhawan, Mirabel \& Rodriguez 2000; Klein-Wolt et
al. 2002). Whilst displaying a highly complex pattern of X-ray
behaviour (e.g. Belloni et al. 2000), the source does at times exhibit
recognisable patterns. For example, it seems that the system
regularly, perhaps always, follows extended hard X-ray periods
(`plateaux') with major ejections (Fender et al. 1999; Klein-Wolt et
al. 2002). These plateaux are themselves associated with powerful
compact jets (Dhawan et al. 2000b; ) which have spectral similarities
with self-absorbed jets from other black hole candidate (BHC) systems
in similar hard X-ray states (Fender 2001). Furthermore, the
observation of apparent superluminal motions from several other XRB
BHCs (e.g. Mirabel \& Rodriguez 1999) and the fairly typical `radio
loudness' (Fender \& Kuulkers 2001) imply that the jets of GRS
1915+105 may be far more typical of those from XRBs than the
relatively slow, cool, jets of SS 433.

In this paper we report the discovery of a strong and variable
circularly polarised component in the radio emission from GRS 1915+105
which we can directly associate with relativistic ejection events.

\section{Observations}

In Fig 1 we show radio and soft X-ray monitoring of GRS
1915+105, over a 150-day period. The radio monitoring data were
obtained with the Ryle Telescope (RT), at a frequency of 15 GHz; for a
more detailed description of this monitoring program see Pooley \&
Fender (1997). The X-ray data are from the {\em Rossi}XTE All-Sky
Monitor (ASM) and measure the total flux in the 2-12 keV band. The
{\em Rossi}XTE ASM is described in Levine et al. (1996) and the public
data can be obtained at {\bf xte.mit.edu}.

Indicated in the top panels of Fig 1 are the times of our two ATCA and
multiple MERLIN observations of GRS 1915+105.

\begin{figure}
\centerline{\epsfig{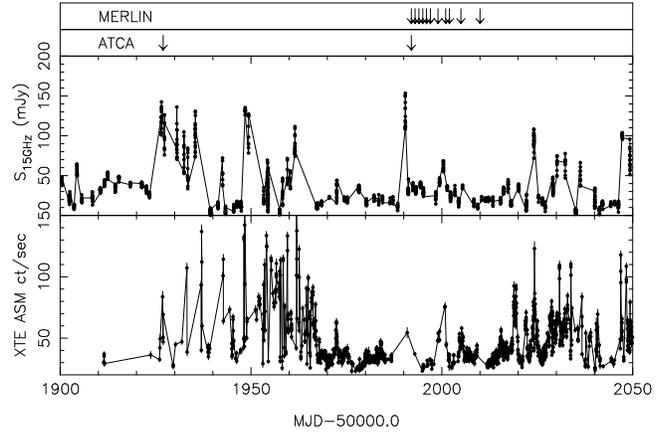}}
\caption{Daily radio and X-ray monitoring of GRS 1915+105, with the
times of our ATCA and MERLIN observations indicated.}
\end{figure}

\subsection{ATCA}

The Australia Telescope Compact Array (ATCA; Frater, Brooks \& Whiteoak
1992) has a number of design features which enable very accurate
circular polarization measurements. The low antenna cross-polarization
and high polarization stability enable accurate calibration of
polarization leakage terms, and the linearly-polarized feed design
largely isolates Stokes V from contamination by Stokes I.

ATCA observed GRS 1915+105 twice, for six hours each, on 2001 January
17 and 2001 March 23. During the January observations, simultaneous
observations at 1384 MHz and 2496 MHz were interleaved with
observations at 4800 MHz and 8640 MHz; for the March observations,
only 4800 MHz and 8640 MHz were observed. For both epochs the array
was in a `6 km' configuration, for which the lack of short baselines
served to reduce confusion from other galactic sources. The
observation and calibration procedures were similar to those described
in Fender et al. (2000).

\begin{figure*}
{\epsfig{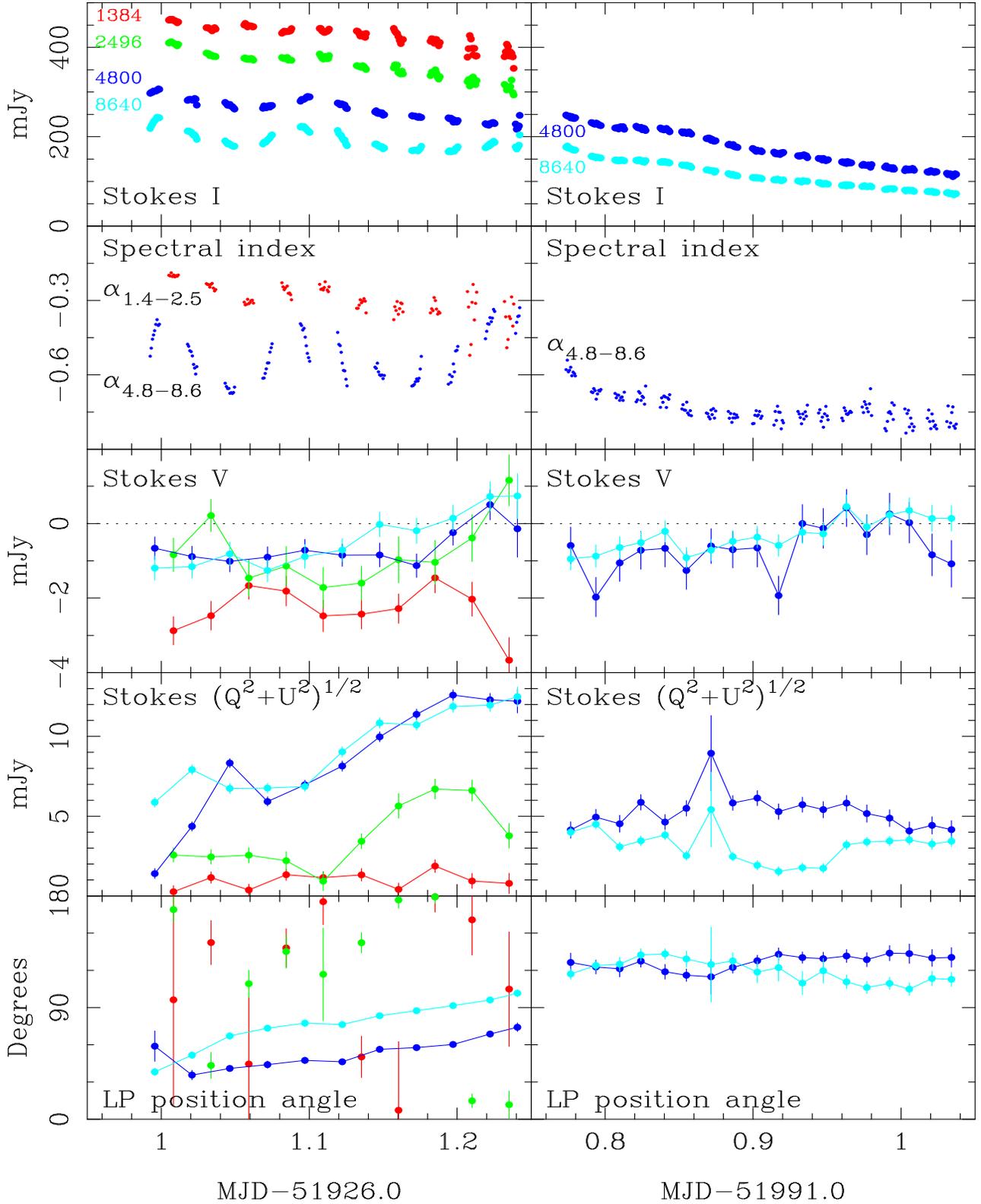}}
\caption{Dual-frequency full polarisation measurements of GRS 1915+105
in 2001 January (left) and March (right). Circular polarisation is
detected at all frequencies observed. Note also the correlated change
in linear polarisation strength and position angle at 4.8 \& 8.6 GHz in
2001 January, a `rotator' event.}
\end{figure*}

As discussed in Fender et al. (2000), calibration of circular
polarization data requires the "strongly-polarized" calibration
equations (Sault, Killeen \& Kesteven, 1991), using a point-source with
a few percent linear polarization. This is needed to calibrate the
leakage of linear polarization into circular. For the 4800 MHz and
8640MHz observations, the VLA calibrator 1923+210 was used as a
polarization calibrator for both epochs. Calibrator confusion and low
linear polarization, however, precluded the use of any of the observed
calibrators as polarization calibrators for the January 1384 MHz and
2496 MHz observations. As a result, we were forced to use calibration
solutions derived using the "weakly-polarized" equations with the ATCA
primary calibrator, 1934-638.

The use of the "weakly-polarized" equations will cause a time-varying
leakage of linear polarization into circular. In tests, peak leakages
of 5\% of the linear polarization into circular have been observed.
For the 1384 MHz observations, the low linear polarization of GRS
1915+105 implies the effect of such leakage is negligible. Even for
the 2496 MHz observations, where the linear polarization rises
rapidly during the observation, in the {\em worst-case} the leakage
would be only half the Stokes V error due to thermal noise. The full
polarisation ATCA data for both epochs are presented in Fig 2.

\subsection{MERLIN}

The Multi Element Radio Linked Interferometer Network (MERLIN)
consists of six individual antennae with a typical diameter of 25m and
a maximum baseline of 217 km (Thomasson 1986). The observations
presented here were undertaken in continuum mode at a frequency of
4994 MHz with a total bandwidth of 16 MHz.  As MERLIN measures all
four correlation products as a matter of course when in this mode,
full polarimetric information can be derived from all images. Ongoing
work is seeking to establish the reliability of Stokes V measurements
with MERLIN; these will not be reported here.

GRS 1915+105 was observed eleven times with MERLIN following the flare
observed on 2001, March 22/23. The first five epochs, corresponding to
daily observations between 2001 March 24 and March 27 and again on
March 29, are presented in this paper (Fig 3); further details and
analysis of the full set of MERLIN observations will be published in
McCormick et al. (in prep). In each case a flux calibrator, 3C286, a
point source, OQ208, and a phase calibrator, 1919+086 , were included
in the observing schedule. The flux calibrator and point source
calibrator were observed at the beginning and end of the run whilst
the rest of the observation was devoted to a cycle of 1.5 minutes on
the phase calibrator and 5 minutes on GRS 1915+105.

Initial data editing and calibration were performed using the standard
MERLIN d-programs and the data were then transferred to the NRAO
Astronomical Image Processing System (AIPS).  Within AIPS the data
were processed via the MERLIN pipeline, which calibrates and images
the phase reference source and then applies these solutions to the
target source. This process also derives instrumental polarisation
corrections and calibrates the linear polarisation position angle,
using 3C286 as the calibrator and assuming a position angle of
33\degr for its E vector. The position angles measured by MERLIN and
ATCA are consistent with the same value, independently confirming the
position angle calibration of each array.

Further self calibration was then carried out within AIPS and GRS
1915+105 imaged in total intensity and stokes Q and U. These maps were
then combined using the AIPS task PCNTR to produce the final maps with
total intensity contours and vectors denoting the direction and
strength of linear polarisation.

Note that we can be confident both from previous studies (e.g. Mirabel
\& Rodriguez 1994; Fender et al. 1999) and these data (McCormick et
al. in prep) that the component(s) to the south east (labelled in Fig
3 as `SE1') is `approaching', and component(s) to the north west are
`receding' (although in fact both sides of the jet have Doppler
factors $\delta < 1$).

\begin{figure}
\centerline{\epsfig{file=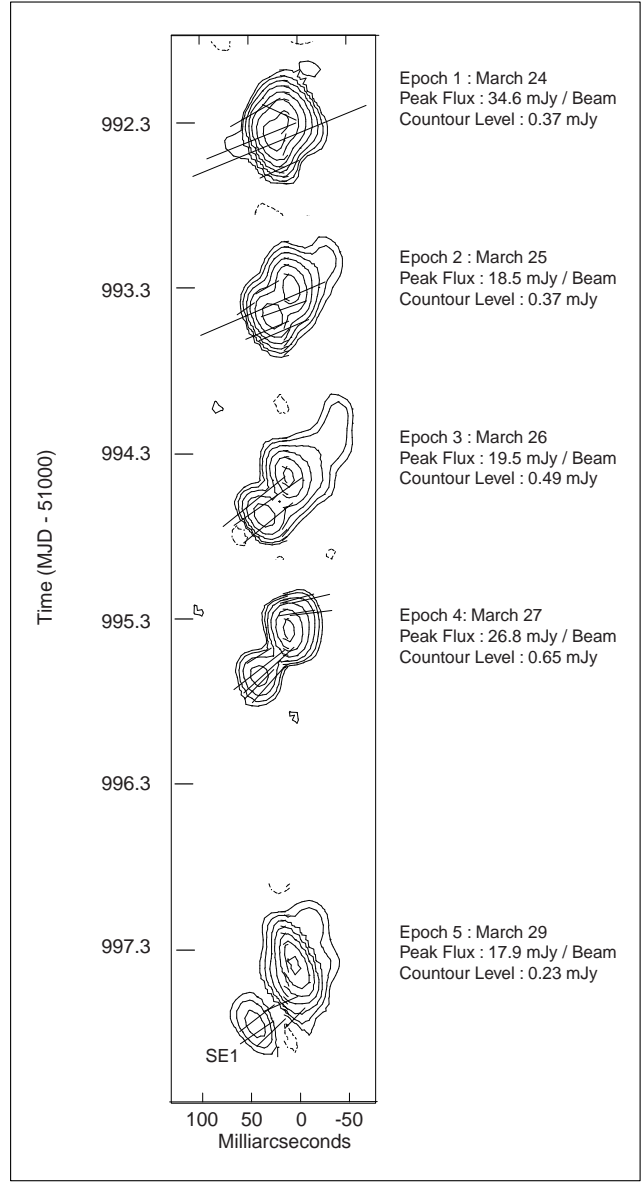,angle=0,width=8.4cm,clip=}}
\caption{MERLIN imaging of relativistic (superluminal) ejections from
GRS 1915+105. We can confidently associated most of the flux measured
by ATCA (Fig 2) with the ejected component to the SE of the core.
Contours are at -1,1,2,4,8,16,32,64 times the level listed by each image.
}
\end{figure}

\section{Variable circular polarisation}

In both sets of ATCA observations, GRS 1915+105 is unambiguously
detected as a source of circularly polarised radio emission (Stokes
V).  


\begin{figure*}
\centerline{\epsfig{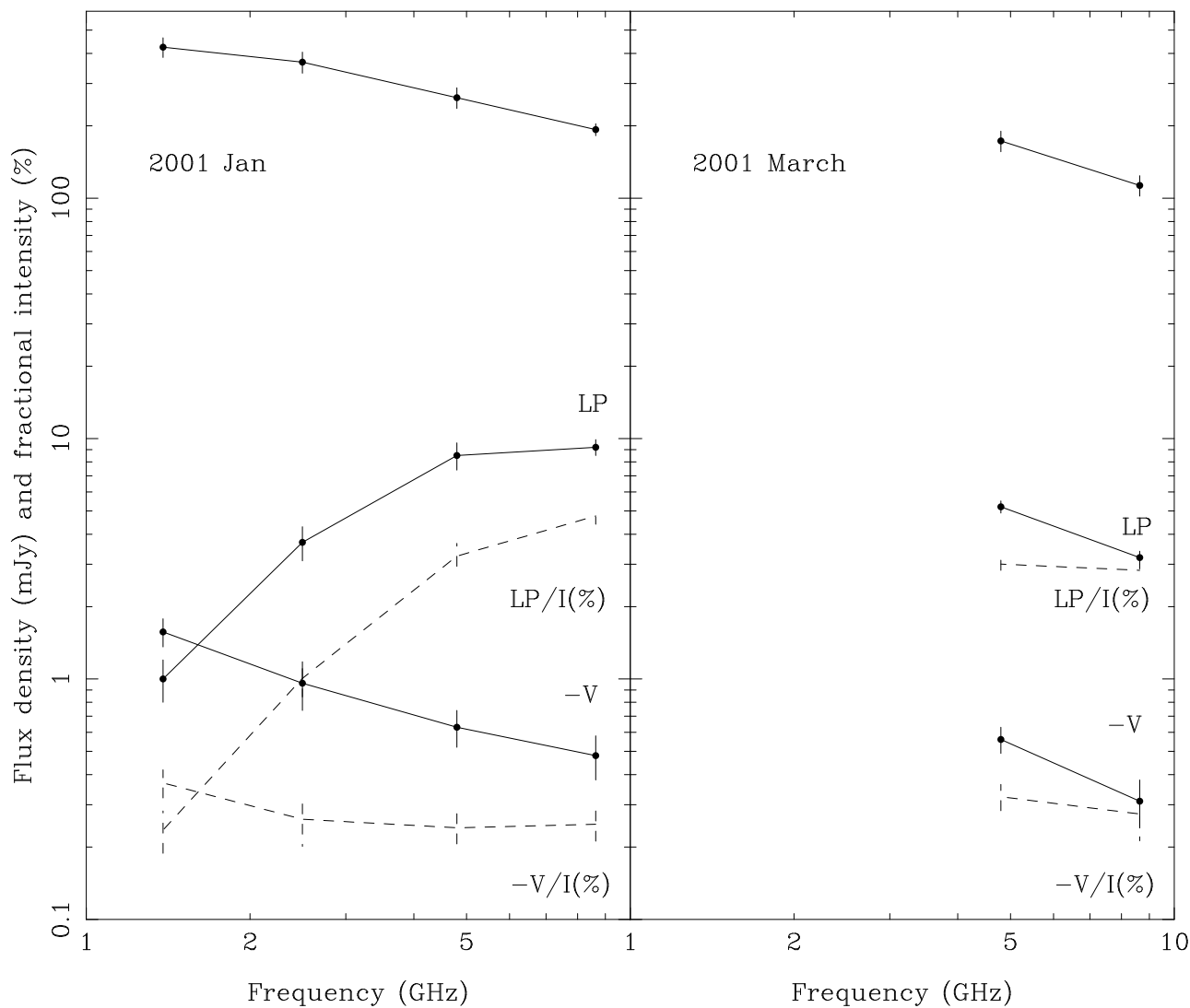}}
\caption{
Mean total flux density, linear and circular polarisation spectra for
GRS 1915+105 in 2001 January and March, from the data in table
1. Note, in 2001 January, the inversion in the linear to circular
polarisation ratio between the two lowest frequencies.}
\end{figure*}

\subsection{2001 January 17}

In 2001 January (Fig 2, left panels), significant CP is measured at
all four ATCA frequencies, from 1--9 GHz. The total flux density is
clearly declining, indicating the decay phase of a major flare, but
there is also significant variability superposed on the relatively
smooth decline, preferentially at higher frequencies. This is almost
certainly indicative of repeated activity in the core, corresponding
to fresh ejection events. The spectral indices support this
interpretation; between 1.4--2.9 GHz the spectrum is significantly
flatter than expected for optically thin synchrotron emission; between
4.8--8.6 GHz it is displaying the rapidly varying behaviour associated
with `core' ejection events (Fender et al. 2002).

Inspection of the total flux and spectral index light curves indicates
there were at least four separate ejection events contributing to the
light curve at this epoch. Fig 1 also indicates that this outburst was
more prolonged than that in 2001 March (see below).

The CP flux is clearly rising to lower frequencies, but the exact
fractional spectrum is difficult to determine as the
multiple components contributing to the observed emission are
unresolved with ATCA. Table 1 lists the mean total, linearly polarised
and circularly polarised flux densities, and Fig 4 plots these both as
total and fractional spectra. We also note that there are measurements
when the Stokes V flux is not significantly non-zero, and even a  few
points where it appears to have changed sign. However, (i) the mean
Stokes V fluxes are significant, and negative (at both epochs), (ii)
the apparent Stokes V sign change has a significance $<2\sigma$ and so
we do not consider it convincing.

\begin{centering}
\begin{table*}\label{march01}
\caption{Mean total flux densities, linear and circular polarisations
of GRS 1915+105 as measured by ATCA, 2001 January and March. The
spectra resulting from these data are plotted in Fig 4.}
\begin{tabular}{|rcccccc|}
\hline
 & \multicolumn{3}{c}{2001 January} & \multicolumn{3}{c}{2001 March}\\
$\nu$ (MHz) & I (mJy) & LP (mJy) & V (mJy) & I (mJy) & LP (mJy) & V
(mJy) \\
\hline
1384& $425 \pm 40$ & $1.0\pm 0.2$ & $-1.57\pm 0.21$ & && \\ 
2496& $368 \pm 37$ & $3.7 \pm 0.6$ & $-0.96\pm 0.22$ &&&\\ 
4800&  $262 \pm 26$ & $8.5 \pm 1.1$ & $-0.63 \pm 0.11$ & $173 \pm 17$
& $5.2 \pm 0.3$ & $-0.56 \pm 0.07$\\
8640&  $193 \pm 11$&  $9.2 \pm 0.7$&  $-0.48 \pm 0.10$ &$113\pm 11$ &
$3.2\pm 0.2$ & $-0.31 \pm 0.07$  \\ 
\hline
\end{tabular}
\end{table*}
\end{centering}

\subsection{2001 March 23}

In 2001 March, GRS 1915+105 was again observed to flare in our 15 GHz
monitoring program, reaching $\sim 160$ mJy at MJD 51990.43.
This time we triggered both ATCA and MERLIN -- in fact the first epoch
of MERLIN observations started at almost exactly the same time as the
ATCA run concluded (Figs 3,5). As a result, we were able to
definitively associate the outburst with relativistic ejections from
the system (Fig 3).

The full polarisation ATCA data are presented in Fig 2 (right panels),
and it is clear that there is less variability in the light curve than
in 2001 January, with the smooth decay in radio flux at both
frequencies only interrupted by the temporary increase around MJD
51991.82. Assuming the emission observed is associated with the radio
event on which we triggered, our ATCA observations commence
$\sim 1.4$ days after ejection (assuming a Doppler factor of $\sim
0.3$ -- Fender et al. 1999 -- this corresponds to only $\sim 10$ hr of
evolution in the rest frame of the ejecta). The 4.8--8.6 GHz spectral
index clearly demonstrates that the majority of the emission is coming
from optically thin regions.  The mean total intensity, linearly
polarised and circularly polarised flux densities are presented in
Table 1 and Fig 4.

\begin{figure}
\centerline{\epsfig{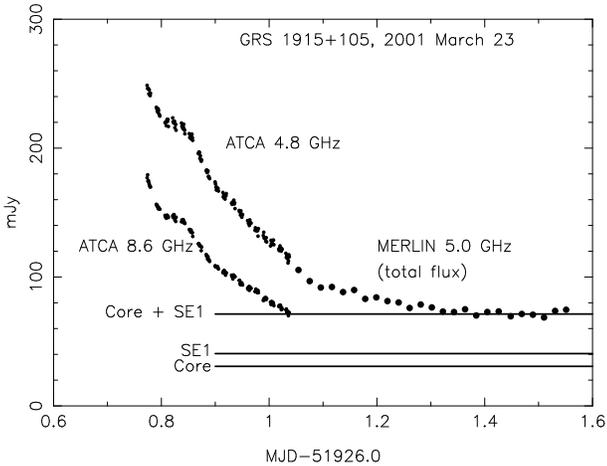}}
\caption{ATCA 4800 \& 8640 MHz light curves compared with exactly
adjacent MERLIN 5000 MHz total intensity light curves on 
March 23, 2001. Also indicated are the levels of the core and SE1
(ejecta) components as measured from the MERLIN image at that epoch.
From combination
of this and the images in Fig 3 we can be confident that ATCA measured
mostly flux from the ejected component.} 
\end{figure}

\section{A linear polarisation `rotator' event}

In the lower two panels of Fig 2 the linearly polarised flux densities
and electric vector position angles are indicated. It is quite evident
from the lower panel that over the $\sim 6$ hr of the observation in
January 2001, both 4800 and 8640 MHz electric vectors rotate smoothly
through $\sim 50$ degrees. Note that while there is evidence for
changing absorption on the scales over which the ejecta from GRS
1915+105 can be tracked (Dhawan, Goss \& Rodriguez 2000), variable
Faraday rotation cannot be responsible for this effect since the {\em
separation} between the vectors at the two frequencies remains
constant (if Faraday rotation, which varies as $\lambda^2$, a rotation
of $\sim 50$ degrees at 8.6 GHz should have a corresponding rotation
of $\sim 100$ degrees at 4.8 GHz).

The smooth rotation in the electric vector position angle seems to be
a little at odds with both the MERLIN observations of Fender et
al. (1999) in which the electric vector was varying seemingly
erratically from day-to-day, and the observations presented here (both
ATCA and MERLIN) for 2001 March, in which the vector remains
approximately constant. Such smooth rotation indicates either a
genuinely rotating jet or, perhaps more likely, a smooth change in the
(projected) position angle of the magnetic field in the emitting
region -- such as a global curved structure in the jet.

Similar behaviour, `polarisation rotator events' -- see Saikia \&
Salter (1988) and references therein - has also been observed in
AGN. While initially interpreted as physical rotation of the magnetic
field structure (which could directly link a jet to e.g. removal of
angular momentum from an accretion flow) the more favoured
interpretation is the formation of a shock inclined at some angle to
the line of sight. The lack of the repeat of this phenomenon in 2001
March would seem to indicate it does not reflect physical rotation of
the jet, which we would assume either always happens or never happens.
However, more recently Gomez et al. (2001) have interpreted the steady
rotation of the linear polarisation vector of a superluminal component
in the jet of the AGN 3C120 as indicating an underlying twisted
(helical) magnetic field structure.

\section{Discussion}

The origin of a circularly polarised component in the radio emission
from AGN and X-ray binaries remains uncertain. However, in both
classes of object the bulk of the radio emission can be confidently
assumed to arise from similar physical processes, namely synchrotron
emission from relativistic electrons in a magnetised plasma flowing
away from the central black hole (or neutron star) in collimated
jets. Since we have every reason to believe that the Stokes I, and
probably Q and U, fluxes from these objects arise via the same
processes, we can hope that by studying the origin of Stokes V in
X-ray binaries we may shed light on its origin in AGN, and vice versa.

\subsection{Association of CP with young ejections}

What can we learn from these observations of GRS 1915+105 ?  From
analysis of the combined ATCA and MERLIN data sets for 2001 March, we
are confident that the measured circularly polarised radio emission is
associated with the relativistic ejection SE1. Our reasoning is as
follows:

\begin{enumerate}
\item{The decreasing Stokes I flux measured at both epochs 
(most obviously 2001 March)
arises from
ejected components whose radio flux decays steadily, probably due to
adiabatic expansion losses, with time. This can be inferred both from
past experience and directly from the MERLIN observations which
directly image the fading ejecta as they propagate away from the
core. In Fig 5 we show the ATCA light curves, plus the total flux
light curve from MERLIN and the fluxes of the two components (core and
ejection SE1). The core stays at roughly the same flux level over all
of the first five epochs of MERLIN imaging, whereas the flux of SE1
continues to fade -- from this we can infer that the bright and
decreasing flux observed by ATCA is dominated by emission from ejected
component SE1.}

\item{There is a significant correlation between the Stokes I and
(-)Stokes V fluxes, especially at 8640 MHz. In table 2 we list the
Spearman rank correlation coefficients between the Stokes I and V
fluxes at 4800 and 8640 MHz, for both the 2001 January and March data
sets; these are plotted in Fig 6. For all epochs, and the combined
data sets, there is a significant rank correlation between Stokes I
and V at 8640 MHz; at 4800 MHz the correlation is marginal. Since the
Stokes I, as argued above, is associated with SE1, we can therefore be
confident that the Stokes V flux also arises primarily in SE1. Note
that we cannot rule out a Stokes V flux of amplitude $|V| \la 0.2$ mJy
associated with the core, based on Fig 6.}

\item{Furthermore, the MERLIN imaging clearly shows that the linearly
polarised radio emission also arises in component SE1. Therefore we
are able to accurately measure the fractional linear and circular
polarisation (ie. all Stokes parameters) for the synchrotron emission
from a single optically thin component}
\end{enumerate}

\begin{table*}
\caption{Spearman rank correlation coefficients for Stokes I vs. V}
\begin{tabular}{|cccccccccc|}
\hline
 & \multicolumn{3}{c}{2001 January} & \multicolumn{3}{c}{2001 March} &
\multicolumn{3}{c}{Combined data sets} \\
 & N & $r_s$ & P & N & $r_s$ & P & N & $r_s$ & P \\
\hline
4800 MHz & 11 & -0.34 & $\leq 70$\% & 18 & -0.42 & $>90$\% & 29 &
-0.32 & $\sim 90$\% \\
8640 MHz & 11 & -0.76 & $>98$\% & 18 & -0.84 & $>99$\% & 29 & -0.52 &
$>99$\% \\
\hline
\end{tabular}
\end{table*}

\begin{figure}
\centerline{\epsfig{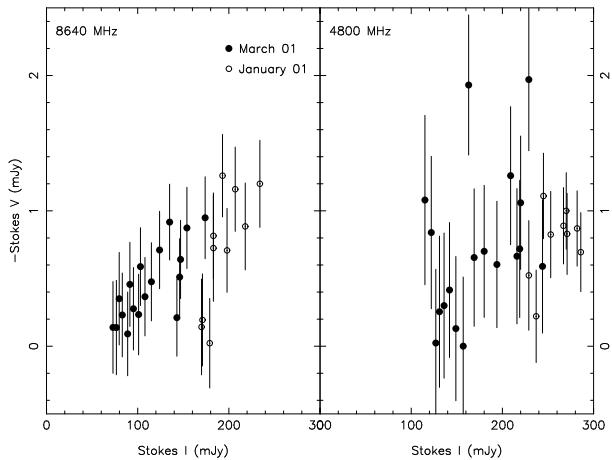}}
\caption{Correlation between Stokes I and V flux densities. Spearman
rank correlation coefficients are listed in table 2. The 8640 MHz
Stokes I and V are (rank) correlated at $>99$\% confidence at both
epochs, and as a combined data set. The 4800 MHz data are marginally
correlated at the $\leq 70$\% confidence level in 2001 January, rising
to $\sim 90$\% confidence in 2001 March and for both data sets combined.}
\end{figure}

\subsection{Comparison with AGN, Sgr A* and M81*}

Rayner et al. (2000) and Homan et al. (2001) have established that
most AGN have ratios of linear to (absolute) circular polarisation
$\geq 10$, whereas the low-luminosity radio cores Sgr A* and M81* have
ratios $\leq 1$ (Bower et al. 1999; Sault \& Macquart 1999; Brunthaler
et al. 2001). It is interesting that for the two XRBs for which we
have so far measured CP, we find both situations, depending on the
frequency observed. In both SS 433 (Fender et al. 2000) and GRS
1915+105, LP $<$ CP at the lowest frequency (1.4 GHz), and LP $\ga$ CP
at higher ($\geq 2.5$ GHz) frequencies. In GRS 1915+105, it is clear
from Figs 2,4 that there is significant opacity (foreground or
internal) at the lower frequencies, and so we consider it most likely
that the reduction in LP is due to Faraday depolarisation.  While in
SS 433 there is less obvious indication of opacity at the lowest
frequencies, it would seem likely that the same mechanism is operating
there to reduce the observed LP. Furthermore, M81* (Brunthaler et
al. 2001) and a number of AGN (Rayner 2000) also seem to show
flat/inverted fractional CP spectra, as we seem to have measured from
GRS 1915+105 in 2001 January.  However, we should also note that in
Sgr A* there is no optically thin power law component, unlike the two
X-ray binaries discussed here, which will further reduce the expected
LP (Bower et al. 1999).

In addition, the majority of AGN are found to maintain the same sign
of Stokes V on timescales from months to decades (Komesaroff et
al. 1984; Rayner et al. 2000; Homan et al. 2001). Bower et al. (2002)
have more recently shown that Sgr A* has maintained the same sign (and
level) of CP over 20 years.  In comparison, for X-ray binaries there
are two observations of SS 433 separated by 10 days (Fender et
al. 2000), and two observations of GRS 1915+105 separated by 65 days
presented here; in both cases the sign remains the same. Of course the
X-ray binary sample is extremely small, but should expand rapidly in
the near future, with several events per year bright enough to mean CP
at the $<0.1$\% level (Fender \& Kuulkers 2001). It is interesting to
note that {\em if} accretion timescales scale linearly with mass from
X-ray binaries to AGN (e.g. Sams, Eckart \& Sunyaev 1996) then
timescales of tens of days for SS 433 and GRS 1915+105 would
correspond to timescales of thousands of years for Sgr A* and millions
of years for some AGN -- ie. we may have already probed longer in
`accretion time' than in all the studies of AGN to date !

\subsection{Origins of the CP component}

The CP spectrum detected from GRS 1915+105 is rather similar to that
observed from SS 433 (Fender et al. 2000), being observed over a broad
range (1--9 GHz) and with a decreasing Stokes V (although not
necessarily V/I) spectrum. The broadband nature of the CP spectrum
suggests that coherent emission mechanisms are unlikely. Furthermore,
we do not consider the birefringent scintillation mechanism of
Macquart \& Melrose (2000) very likely either, since the CP component
seems to be associated with a physical event in the source, yet
because of the high velocities of the ejecta it is likely to be a
large distance from the source during the periods in which we measured
CP. 

This leads us (once again) to consider one of two mechanisms most
viable, an intrinsic CP component to the synchrotron emission or
LP$\rightarrow$CP conversion (`repolarisation'). Do we have any
evidence in favour of either of these ? The intrinsic synchrotron
mechanism should, naively, produce a well-defined $\nu^{-0.5}$ V/I
spectrum in a homogenous, optically thin, source. Our observations in
2001 March match these criteria quite closely (supported by direct
imaging of a single ejection event with MERLIN), but unfortunately we
only have a two-point CP spectrum at this epoch. In addition, the
relatively low signal-to-noise ratio of the CP detections only allows
us to constrain the (-V)/I spectral index to be $\alpha_{\rm -V/I} =
-0.3 \pm 0.3$.  In 2001 January the situation is rather more complex,
the mean flux and polarisation spectra certainly containing the
contributions of multiple components with different optical
depths. The data may suggest that the CP arises preferentially in
`core' components with the highest densities and optical depths,
similar to the situation in AGN. This interpretation may favour
instead the LP$\rightarrow$CP `repolarisation' mechanism, which will
operate most efficiently at higher optical depths, but currently the
data are not sufficiently constraining.

However, both mechanisms require a significant population of
low-energy electrons. In principle, a strong probe of this requirement
could be obtained by measuring the low-frequency extent of radio
emission during outbursts of GRS 1915+105 and other systems. Prompt
observations at MHz frequencies could probe the electron distributions
at Lorentz factors of 30 and lower (based on calculations in Fender et
al. 1999) although, depending on the energy and magnetic field density
in the ejecta, this emission may be self-absorbed.  The consequences
for the energetics of ejection events would be significant, especially
in the case of a neutral baryonic plasma with one proton associated
with each electron.

\section{Conclusions}

Following two high-frequency radio flares detected in our 15 GHz
monitoring program, we have triggered sensitive observations with ATCA
in order to measure the CP component from the jet source GRS
1915+105. On both epochs, 65 days apart, we have measured a varying CP
component, with the same sign.  At the first epoch we obtained a
four-frequency spectrum, but lack of spatial information and clear
variability, in particular at higher frequencies (presumably
indicating generation of new components) precluded a confident
determination of the CP spectrum.

At the second epoch, we have less CP coverage (only two frequencies
instead of four) but the ATCA observations are complemented by
high-spatial-resolution MERLIN imaging. From these MERLIN images we
can be confident that the ATCA flux was dominated by the approaching
component of a discrete ejection event, which was moving at a 
relativistic velocity. The fractional linear polarisation of this
ejection is $\sim3$\% at both frequencies, and the fractional circular
polarisation is $\sim 0.3$\% at both frequencies. Furthermore, the
Stokes I spectrum clearly indicates that the source was optically thin
and probably decaying in flux due to adiabatic expansion losses
(ie. there is no evidence of the spectral steepening which would occur
as a result of radiative losses). These are therefore excellent data
against which to test detailed models of synchrotron emission and
plasma composition.

The data are consistent with production of the circularly polarised
component in the earliest phases of new relativistic ejection events,
as seen in AGN. The observations are not consistent with coherent
emission mechanisms or the birefringent scintillation mechanism of
Macquart \& Melrose (2000), but are more likely to arise either
directly from the synchrotron mechanism or by conversion of linear to
circular polarisation in the emitting plasma. Both of these mechanisms
imply a low-energy tail to the electron distribution which would have
a significant impact on our estimates of the energy associated with
such relativistic ejections. Our data may favour the LP$\rightarrow$CP
conversion mechanism, but are not of sufficiently high quality to be
conclusive. Note that Macquart (2002) reports the detection of
circularly polarised radio emission from another X-ray binary jet
source, GRO J1655-40, so this may well turn out to be a common feature
and important diagnostic of the jets from X-ray binaries.

Finally, smoothly rotating 4800 and 8640 MHz electric vectors and
correlated LP flux densities in Jan 2001 indicate a changing
projection of the magnetic field, possibly indicating genuine rotation
of the jet/field structure or the progressive formation of a shock in
the outflow.

\section*{Acknowledgements}

We would like to thank the referee, Heino Falcke, for useful comments.
The Australia Telescope is funded by the Commonwealth of Australia for
operation as a National Facility managed by CSIRO.  MERLIN is operated
as a National Facility by the Jodrell Bank Observatory of the
University of Manchester, on behalf of the Particle Physics and
Astronomy Research Council (PPARC).  RXTE ASM results were provided by
the ASM/RXTE teams at MIT and at the RXTE SOF and GOF at NASA's
GSFC. We thank the staff at MRAO for maintenance and operation of the
Ryle Telescope, which is supported by the PPARC.

\end{document}